\documentclass[twocolumn,pra,aps,showpacs, flushbottom]{revtex4-1}
\usepackage{xspace,amsmath,amsfonts,amsthm,amssymb,amsbsy,graphicx,color}

\newcommand{\ms}[1]{\mbox{\scriptsize #1}}

\begin{document}

\title{Comparing resolved-sideband cooling and measurement-based feedback cooling \\ on an equal footing: analytical results in the regime of ground-state cooling}


\author{Kurt Jacobs$^{1,2}$, Hendra I. Nurdin$^3$, Frederick W. Strauch$^4$, and Matthew James$^5$} 

\affiliation{$^1$Department of Physics, University of Massachusetts at Boston,
Boston, MA 02125, USA \\
$^2$Hearne Institute for Theoretical Physics, Louisiana State University, 
Baton Rouge, LA 70803, USA\\
$^3$ School of Electrical Engineering and Telecommunications, University of New South Wales, Sydney, NSW 2052, Australia \\
$^4$Department of Physics, Williams College, Williamstown, MA 01267 \\
$^5$Centre for Quantum Computation and Communication Technology, 
School of Engineering, Australian National University, Canberra, ACT 0200, Australia
}

\begin{abstract} 
We show that in the regime of ground-state cooling, simple expressions can be derived for the performance of resolved-sideband cooling --- an example of coherent feedback control --  and optimal linear measurement-based feedback cooling for a harmonic oscillator. These results are valid to leading order in the small parameters that define this regime. They provide insight into the origins of the limitations of coherent and measurement-based feedback for linear systems, and the relationship between them. These limitations are not fundamental bounds imposed by quantum mechanics, but are due to the fact that both cooling methods are restricted to use only a linear interaction with the resonator. We compare the performance of the two methods on an equal footing --- that is, for the same interaction strength --- and confirm that coherent feedback is able to make much better use of the linear interaction than measurement-based feedback. We find that this performance gap is caused not by the back-action noise of the measurement but by the projection noise. We also obtain simple expressions for the maximal cooling that can be obtained by both methods in this regime, optimized over the interaction strength. 
\end{abstract}


\maketitle 

\section{Introduction} 

Preparing mechanical harmonic oscillators in their ground states is potentially important for future quantum technologies~\cite{Poot12}, and is presently relevant for experimental work in optomechanics~\cite{Arcizet06, Gigan06, Schliesser08, Thompson08, Groeblacher09, Eichenfield09b, Tsang10} and nano-electromechanics~\cite{LaHaye09, Massel11, Teufel11, Teufel11b, Palomaki13}. Here we consider two simple and rather different methods for achieving this goal. The first, called resolved-sideband cooling, is an example of coherent feedback control~\cite{James08b, Nurdin09, Hamerly12, Kerckhoff13, Jacobs14b} in which the mechanical oscillator is coupled linearly to an ``auxiliary'' microwave or optical mode~\cite{Marquardt07, Wilson-Rae07, Wang11, Jacobs11c, Machnes12}. Since the auxiliary oscillator has a much higher frequency than the mechanics, it is in its ground state at the ambient temperature. Because of this the coupling between the two transfers both energy and entropy from the mechanics to the auxiliary, cooling the former. Sideband cooling has already allowed experimentalists to prepare mechanical oscillators in a state with less than one phonon. The second method we investigate is that in which an explicit continuous measurement is made on the mechanical oscillator (from now on just ``the oscillator''), and the information from this measurement is used to apply a force to the oscillator to damp its motion in the manner of traditional feedback control~\cite{KSeq}. Our motivation for comparing the performance of resolved-sideband cooling and this measurement-based feedback cooling is to determine how differently the two forms of feedback behave, in the regime of good ground-state cooling, and to understand better the origin of this difference.  

Two previous works have examined, and to varying extents compared, the two cooling methods we consider here. To explain how our work extends and complements these previous analyses we now summarize them briefly. The work by Genes \textit{et al.}~\cite{Genes08} was the first to obtain a complete analytical solution for resolved-sideband cooling. In addition to presenting this solution they also analyzed a measurement-based feedback protocol for cooling in which the raw signal from a continuous measurement of position is processed by taking its derivative,  and a force applied to the oscillator proportional to this processed signal. Nevertheless, Genes~\textit{et al.} were not able to compare quantitatively the effectiveness of the two methods because they did not have a means to quantitatively compare the resources used by each: sideband cooling employs a unitary coupling to the oscillator, whereas measurement-based feedback employs an irreversible coupling quantified by a damping rate. 

Hamerly and Mabuchi, employing the theory developed in~\cite{Nurdin09, James08b}, made a direct quantitative comparison of the effectiveness of sideband cooling and measurement-based feedback by using the fact that both cooling methods can be realized by coupling the mechanical oscillator to a traveling-wave electromagnetic field (also known as an \textit{output channel}~\cite{Jacobs11c})~\cite{Hamerly12, Hamerly13}. That is, a traveling-wave field can be used to mediate both the continuous measurement used in measurement-based feedback and the unitary coupling of sideband cooling (coherent feedback). Because both cooling methods can be implemented using the same coupling, one can ask which method is able to make the best use of the information obtained by the coupling for a given coupling rate. Hamerly and Mabuchi (HM) also used the optimal estimates of the mean position and momentum in the measurement-based feedback protocol, as we do here, whereas Genes \textit{et al.} did not. For a weakly-damped (high-Q) mechanical oscillator, and for a fixed set of parameters, HM compared measurement-based cooling to coherent feedback as a function of the bath temperature. They found that for weak damping, and for a given set of parameters, coherent feedback was able to cool better than the best linear measurement-based feedback. 

The results of Hamerly and Mabuchi were purely numerical. Technologically the most interesting regime for cooling is that in which the mechanical oscillator has a high $Q$ factor (weak damping), and in which the coupling rate to the controller is strong enough that the control protocol can keep the mechanical oscillator close to its ground state. Here we show that it is possible to obtain simple analytic expressions for the optimal cooling achieved by both control methods to first order in the small parameters that define the regime of high $Q$ and ground-state cooling. For sideband cooling this is achieved merely by expanding the full expression for the performance to second order in these parameters. For measurement-based cooling the equations that determine the performance are non-linear, and can only be solved exactly for zero damping ($Q=\infty$). We obtain analytic expressions for weak damping to first-order in the small parameters by using a perturbative method that expands about this exact solution. Having analytic expressions for both cooling methods sheds light on the origins of the limits of each, the relationship between these limits, and reveals the dependance on the various key parameters. 

The small parameters that define the regime of high $Q$ and ground-state cooling are as follows. We define the regime of ground-state cooling, which is also the regime of ``good control''~\cite{Li09}, as that in which the control method can maintain the average number of phonons in the oscillator, denoted by $\bar{n}$, at a value much less than unity ($\bar{n} \ll 1$). If we define the steady-state probability that the system will be found outside the ground state by $P_{\ms{e}}$, then this is also the regime in which $P_{\ms{e}} \ll 1$. The rate at which energy flows into the oscillator from the environment is given by $\gamma n_T$ where $\gamma$ is the damping rate of the oscillator and $n_T$ is the average number of phonons that the oscillator would have if it were at the ambient temperature $T$. The regime of ground-state cooling requires that the rate at which the control process extracts energy from the oscillator is much greater than $\gamma n_T$. For coherent feedback this means that the rate of the interaction with the auxiliary, $\lambda$, (defined precisely below), and the damping rate of the auxiliary, $\kappa$, satisfy 
\begin{equation} 
    \lambda \sim \kappa \gg \gamma n_T . 
\end{equation}
For measurement-based feedback the regime of ground-state cooling requires that the measurement rate, $\tilde{k}$, (a scaled version of the measurement strength, defined in Section~\ref{physimp}), and the damping rate induced by the feedback force, $\Gamma$, satisfy  
\begin{equation}
    \Gamma \gg \tilde{k} \gg \gamma n_T . 
\end{equation}
A further requirement for both methods to provide ground-state cooling is that the rate of the linear coupling, $\lambda$, between the oscillator and the auxiliary mode, or the measurement rate $\tilde{k}$, is slower than the frequency $\omega$ of the oscillator. This stems from the fact that a linear interaction is not the ideal interaction for cooling, and it only works well in the weak-coupling regime. This requirement is not as strict as the above inequalities, however, since a value of $\lambda/\omega$ as low as 5 can be sufficient to achieve optimal cooling~\cite{Genes08}. To obtain our simple expressions we do assume that  
\begin{equation}
   \tilde{k}  \ll \omega   ,  \;\;\;\;\;\;\;     \lambda \ll  \omega , 
\end{equation}
and expand to second order in the small parameters $\tilde{k}/\omega$ and $\lambda/\omega$. The fact that these parameters need not be very small is indicated by the fact that they do not affect the cooling to first order but only to second order. Further, as part of our analysis we derive results that are exact in $\tilde{k}/\omega$ and $\lambda/\omega$; it is only in the small parameters $\gamma n_T/\tilde{k}$ and $\gamma n_T/\omega$ for which our results are necessarily perturbative. 

Given the above time-scale separations, the small parameters that define our regime are 
\begin{equation}
   \varepsilon_1 = \frac{\gamma n_T}{\kappa} , \;\;\; \varepsilon_2 = \frac{\gamma n_T}{\lambda} , \;\;\;  \varepsilon_3 = \frac{\kappa}{\omega}, \;\;\; \varepsilon_4 =\frac{\lambda}{\omega} ,  \label{sbsmallp}
\end{equation} 
for resolved-sideband cooling and 
\begin{equation} 
   \varepsilon_5 = \frac{\gamma n_T}{\tilde{k}} , \;\;\;  \varepsilon_6 = \frac{\tilde{k}}{\omega}, 
\end{equation} 
for measurement-based cooling. The size of the ratio $\tilde{k}/\Gamma$ is determined by further considerations that we discuss below. The ratio  $\gamma n_T/\omega$ is second order in the above parameters: 
\begin{equation}
   \frac{\gamma n_T}{\omega} =  \varepsilon_1 \varepsilon_3 = \varepsilon_2 \varepsilon_4 = \varepsilon_5 \varepsilon_6 . 
\end{equation} 
We obtain analytic expressions for the steady-state average phonon number in the oscillator, $\bar{n}$, either to leading or next-to-leading order in the parameters $\varepsilon_i$.  

Resolved-sideband cooling is traditionally implemented by coupling the mechanical oscillator to the auxiliary directly via the linear interaction 
\begin{equation}
   H_{\ms{int}}^{\ms{SB}} = \hbar \frac{\lambda}{2} \tilde{x} X ,  \label{hint}
\end{equation}
where $\tilde{x} = b + b^\dagger$, with $b$ the oscillator annihilation operator, and $X = a + a^\dagger$ with $a$ the annihilation operator of the optical or superconducting mode. The interaction rate $\lambda$ is modulated at the frequency difference between the oscillator and the cavity mode, which is what allows them to exchange energy as if they were resonant. What enables the direct comparison between the two cooling methods is that resolved-sideband cooling can be implemented by coupling the oscillator and auxiliary via a propagating electromagnetic field, and this is also how measurement-based feedback is implemented. In the latter the propagating field is measured using homodyne detection. Thus both cooling methods are able to use the same interface to the resonator, and thus extract information from the resonator in an identical way. For a given rate, $\tilde{k}$, at which the propagating field couples to the oscillator, we can then ask which cooling method performs better, and is thus able to make better use of the information. When the propagating field is measured, the coupling rate $\tilde{k}$ becomes  the ``measurement strength'' (defined below) characterizing the rate at which the measurement extracts information. When the field is used instead to create the linear coupling $H_{\ms{int}}^{\ms{SB}}$ with the auxiliary oscillator, the resulting  interaction rate $\lambda$ realized by the field coupling rate $\tilde{k}$ is 
\begin{equation} 
    \lambda = \sqrt{8\tilde{k}\kappa} , 
\end{equation}
where $\kappa$ is the damping rate of the auxiliary oscillator. 

It is useful in our analysis below to define a variable $\varepsilon$ that represents an expression that is first order, and only first order in all of the small parameters $\varepsilon_k$ that appear in it. This allows us in what follows to indicate that an expression $E$ is order $d$ in any (or all) of the $\varepsilon_k$ with the notation $E \sim \varepsilon^{d}$. 

In the next section we present the physical implementation of both cooling methods via an irreversible output coupling. In Section~\ref{cohfb} we analyze resolved-sideband cooling and derive the expressions for the performance. To do this we use a slightly different approximate master equation to describe the thermal noise of the oscillator than that used by Genes \textit{et al.}~\cite{Genes08}. Their approximation was valid for all damping rates of the oscillator at high temperature, while ours is valid for weak damping at all temperatures. In Section~\ref{measfb} we derive the expressions for the performance of the (optimal) linear measurement-based feedback cooling. In Section~\ref{seccomp} we compare and discuss the performance of the two methods, and the origin of their respective limitations. In Appendix A we discuss how the measurement-based cooling scheme can be treated using the quantum noise equations of input-output theory. In Appendix B we show how the steady-state for resolved-sideband cooling is obtained by integrating the spectrum using a remarkable integral formula. This formula can be used to integrate the spectra of the coordinates for any linear input-output network. 

\section{Physical implementation of the two cooling schemes}
\label{physimp}

\begin{figure}[t] 
\leavevmode\includegraphics[width=0.9\hsize]{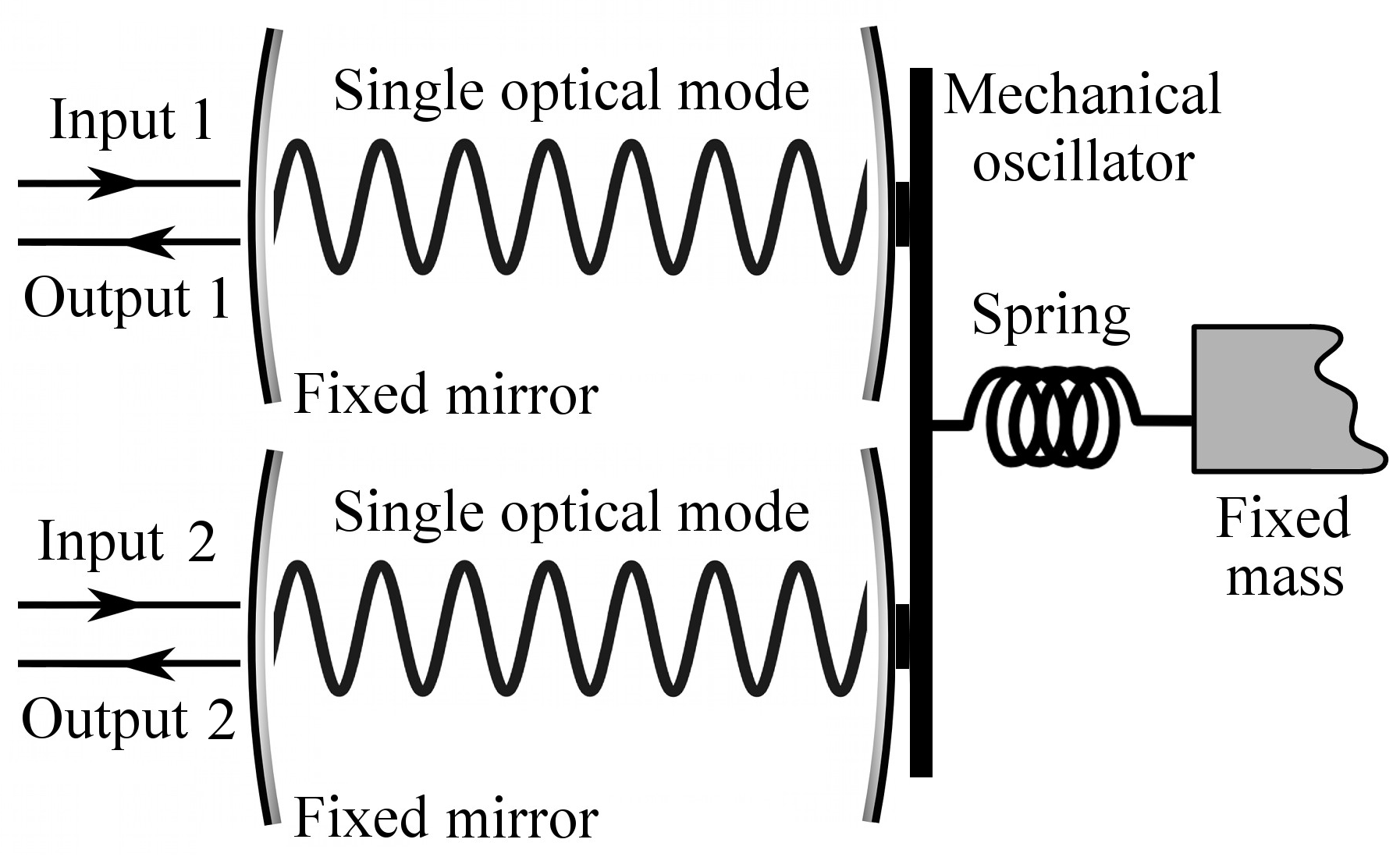} 
\caption{Here we show the mechanical oscillator and the two optical or superconducting cavity modes that provide the interfaces to the resonator that will be used by the controllers. Each cavity provides a coupling to the resonators position. This means that the output of each cavity provides information about the position of the oscillator, and the input of each provides a way to apply a Hamiltonian that is proportional to position, and thus to apply a linear force to the resonator. We have depicted the two cavity modes in separate Fabry-Perot cavities merely for clarity. In an experimental realization the two modes might be, e.g., two counter-propagating modes in a single ring cavity.} 
\label{fig1} 
\end{figure}

The interface via which both cooling methods interact with the oscillator is shown in Fig.~\ref{fig1}. It involves two optical (or superconducting) cavities, each with a single mode, where the right-hand end-mirror of each cavity is attached to the oscillator, and thus oscillates with it. To understand how the interface works, consider the effect of bouncing a beam of light off the oscillator. This beam of light provides both an interface to extract information and to apply a force to the oscillator: i) when the photons in the light beam are reflected from the surface of the oscillator they apply a force to it, and the size of the force is proportional to the beam intensity; ii) monitoring the phase of the reflected light provides a continuous measurement of the position of the oscillator. 

Each of the two optical cavities that are attached to the oscillator provides essentially the same interface as a single beam of light. The photons in the single mode apply a force to the oscillator as they are reflected off it, and the number of photons in the mode can be adjusted by changing the intensity of the light incident on the cavity (e.g. the light entering the top cavity via input 1). The phase of the light that exits each of the cavities provides information about the oscillator position. The faster the light leaks out of the cavities, characterized by their respective damping rates, the more closely each cavity acts like a beam of light reflected from the oscillator.  

While we could use a single optical cavity, with a single mode, to provide both a measurement and a feedback force, we choose to use two cavities because this configuration is required to implement resolved-sideband cooling. For both cooling methods the top cavity will be used as a ``measurement interface'' to extract information, and the bottom cavity will be used as an ``actuation'' interface to apply a force to the oscillator. To compare the two control methods, it is the measurement interface, the interface implemented by the top cavity, that we will demand is the same for both methods. That is, both methods will extract information at the same rate using this interface. As far as the physical implementation is concerned, this means that laser 1 has the same power, $P_1$, and the top cavity the same damping rate, $\gamma_{\ms{tcav}}$, for both methods. 

To use the top cavity to create an interface that provides continuous information about the position of the oscillator, $x$, we set the cavity damping rate, $\gamma_{\ms{tcav}}$, to be much larger than the opto-mechanical coupling rate between the cavity mode and the oscillator, and adiabatically eliminate the cavity. This procedure is detailed in a number of places (e.g.~\cite{Wiseman93, DJ99, Jacobs14}) and we won't repeat it here. The resulting interface can be described by writing the electromagnetic field output from the cavity as~\cite{Hamerly13, Jacobs14} 
\begin{equation}
   c_1^{\ms{out}}(t) = \sqrt{k} x - c_1^{\ms{in}}(t) , 
\end{equation}
where $c_1^{\ms{in}}(t)$ is the input to the cavity. This is the input-output formalism of Collett and Gardiner~\cite{Collett84, Gardiner85}. The constant $k$ characterizes the rate at which the output channel provides information about the position, and is given by $k =  2 \alpha g / (\Delta x \gamma_{\ms{tcav}})$. Here $|\alpha|^2$ is the steady-state number of photons in the cavity, $g$ is the single-photon optomechanical coupling rate~\cite{Mancini97, Bose97}, and $\Delta x = \sqrt{\hbar/(2m\omega)}$ is the ground-state position uncertainty of the mechanical oscillator, with $\omega$ the oscillator frequency and $m$ its mass. The fact that the interface provides information about position, rather than any other observable, and the fixed information rate $k$ are the only limits imposed on our two control protocols. Given this interface, we wish to know which protocol is able to provide the best ground-state cooling, and under what circumstances. 

Here we will use scaled position and momentum variables for the oscillator, $\tilde{x} = b + b^\dagger = x/\Delta x$, and $\tilde{p} = -i(b - b^\dagger) = p/\Delta p$, where $\Delta p = \sqrt{\hbar \omega m/2}$. The correspondingly scaled information rate constant is   
\begin{equation}
  \tilde{k} = (\Delta x^2) k =  2 \alpha \left( \frac{g}{\gamma_{\ms{tcav}}}\right) , 
\end{equation}
allowing us to write the output field as $c_1^{\ms{out}}(t) = \sqrt{\tilde{k}} \tilde{x} - c_1^{\ms{in}}(t)$. The field operators $c_1^{\ms{in}}$ and $c_1^{\ms{out}}$ are continuum versions of annihilation operators. The output field $c_1^{\ms{out}}$ has the same correlation functions as the input field, which are 
\begin{equation}
    \langle c_1^{\ms{in}}(t) c_1^{\ms{in}\dagger}(t+\tau) \rangle = \delta(\tau), \;\;\;\;  \langle c_1^{\ms{in}\dagger}(t) c_1^{\ms{in}}(t+\tau) \rangle = 0. 
\end{equation}
While the interface that provides the information will be the same for both control methods, the interface that provides the feedback force will be used differently in each case. We now describe the two cases in turn. 

\begin{figure*}[t] 
\leavevmode\includegraphics[width=1\hsize]{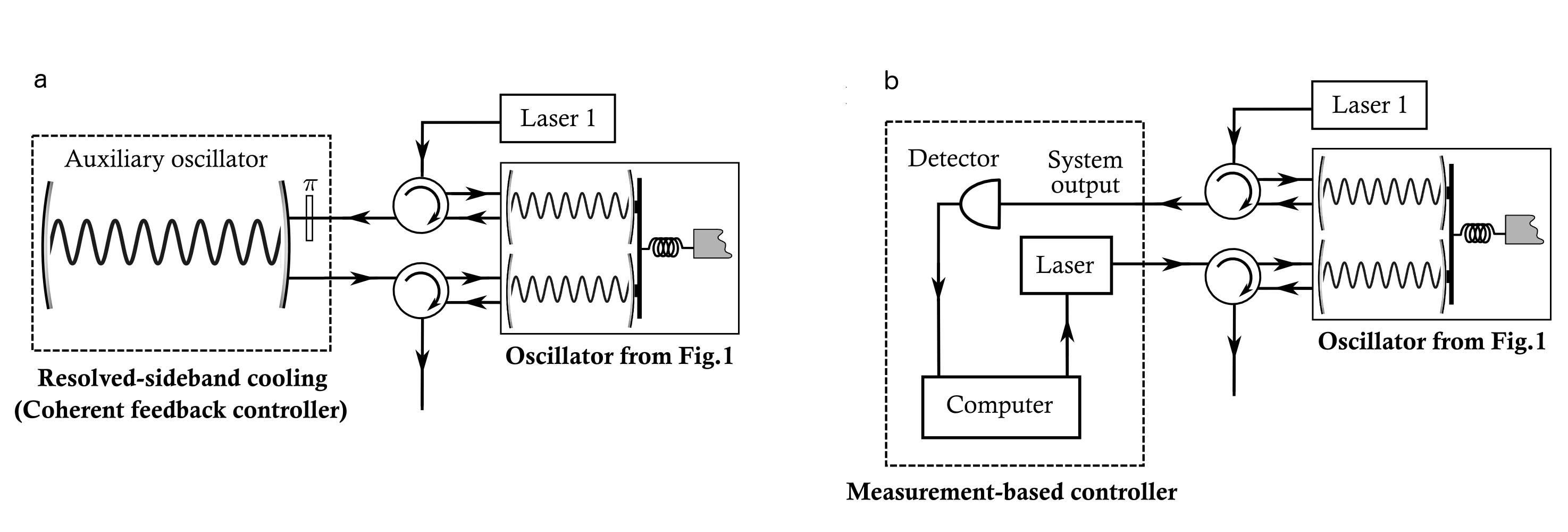} 
\caption{Here we show the configurations that implement the two cooling schemes. The circles are ``circulators'' that separate the outputs of the optical or superconducting cavities from their respective inputs. The two cavities whose right-hand-side end-mirrors are attached to the mechanical oscillator act as interfaces to the oscillator that couple to its position. It is the top interface cavity through which the control system extracts information about the oscillator, and whose output coupling rate, $\tilde{k}$, which is the measurement strength in the case of a continuous measurement, is the same for both schemes. In (b) the detector implements homodyne detection, the computer processes the measurement results and uses the resulting information to control the intensity of the laser.} 
\label{fig2} 
\end{figure*} 

\subsection{Measurement-based feedback}

The configuration that implements measurement-based feedback control is shown in Fig.~\ref{fig2}b. In this case the output field $c_1^{\ms{out}}$ is measured by homodyne detection that monitors the phase of the output light~\cite{Wiseman93, WM10}, and the second interface (cavity 2) is used merely to apply a classical force to the oscillator. To use cavity 2 to apply a classical force we make the damping rate of this cavity, $\gamma_2$, sufficiently large that the information rate provided by output 2 in Fig.~\ref{fig1} goes to zero $(\gamma_2 \gg \alpha_2 g)$. To apply a force to the oscillator we shine a laser into input 2 (see Fig.~\ref{fig2}b) and the resulting force on the oscillator in units of $d\tilde{p}/dt$ is $\tilde{f} = g |\alpha_2|^2 = 4 P_2 /( \hbar\Omega/\gamma_2)$ where $\Omega$ is the frequency of the optical mode in the cavity. We therefore apply a time-dependent force by changing the laser power $P_2$. While it may appear that we can apply only a positive force, this is illusory. The equilibrium position of the oscillator is determined by the force $\tilde{f}$. Thus applying a constant offset force $\tilde{f}_0$, the force on the oscillator with respect to the resulting equilibrium position is $\Delta \tilde{f} = \tilde{f}-\tilde{f_0}$. To peek ahead, the optimal feedback force for the oscillator under linear measurement-based feedback is $\tilde{f}(t) = -\Gamma \langle \tilde{p}(t)\rangle= - \Gamma \mbox{Tr}[\rho \tilde{p}]$, where $\Gamma$ is chosen to be as large as possible. 

The homodyne measurement on output 1 results in a stochastic master equation for the density matrix of the mechanical oscillator describing a measurement with strength $\tilde{k}$. Along with the thermal noise to which the oscillator is subjected, the full master equation describing the dynamics of the oscillator under the measurement-based feedback is~\cite{DJ99, Jacobs14} 
\begin{eqnarray}
   d\rho & = & -i [ \omega b^\dagger b - \tilde{f}(\rho,t) \tilde{x}, \rho ] dt   - \tilde{k} \mathcal{K}(\tilde{x})\rho dt \nonumber \\
    & &  + \sqrt{2 \eta \tilde{k}} \left( \tilde{x}\rho + \rho \tilde{x} - 2\langle \tilde{x}\rangle \rho \right) dW  , \nonumber  \\
                &   &  - \frac{\gamma}{2} (n_T + 1)   \mathcal{K}(b)\rho dt  - \frac{\gamma}{2} n_T \mathcal{K}(b^\dagger)\rho dt , \label{ch8mbfbcx} 
\end{eqnarray} 
where 
\begin{equation}
  \mathcal{K}(c) \rho \equiv c^\dagger c \rho + \rho c^\dagger c - 2 c \rho c^\dagger 
\end{equation}
for an arbitrary operator $c$. Here $0 \leq \eta \leq 1$ is the efficiency of the measurement on output 1, and $n_T$ is the average number of phonons in oscillator at the ambient temperature $T$, given by $n_T = (\exp[\hbar\omega/k_{\ms{B}T}] - 1)^{-1}$, where $k_{\ms{B}}$ is Boltzmann's constant. The damping rate of the oscillator is $\gamma$. As noted above the feedback force is a function of the state $\rho(t)$ at time $t$. 

\subsection{Coherent feedback (resolved-sideband cooling)}

Resolved-sideband cooling is traditionally implemented using a linear interaction between the mechanical resonator and an auxiliary optical or superconducting resonator, as discussed in the introduction. We can use the two interfaces provided by cavities 1 and 2 in Fig.~\ref{fig1} to reproduce this linear interaction. This is done by choosing cavity 2 to have the same parameters as cavity 1, and by applying a $\pi$ phase shift to the light in output 1 (or alternatively input 2), and by connecting the auxiliary optical resonator to output 1 and input 2 as shown in Fig.~\ref{fig2}a. 

In this case it is most convenient to use the quantum Langevin equations of the input-output formalism to describe the dynamics of the auxiliary cavity mode and the oscillator. For the mechanical oscillator these equations are given by 
\begin{equation}
\frac{d}{dt} \left( \!\! \begin{array}{c}  \tilde{x}  \\  \tilde{p}  \end{array} \!\! \right) = 
\left( \!\! \begin{array}{ccc}
- \frac{ \gamma}{2} & \omega    \\
 -\omega  &  - \frac{ \gamma}{2}     
\end{array} \!\! \right) \!\! 
\left( \!\! \begin{array}{c}  \tilde{x}  \\  \tilde{p}  \end{array} \!\! \right) +   \mathbf{v}_{\ms{in}} 
 \label{ddtmeans1}
\end{equation}
with 
\begin{equation}
\mathbf{v}_{\ms{in}} = 
\left( \!\! \begin{array}{l} \sqrt{\gamma}\, x_T^{\ms{in}} \\  \sqrt{\gamma}\, p_T^{\ms{in}} + \sqrt{2\tilde{k}} \, p_1^{\ms{in}} + \sqrt{2\tilde{k}} \, p_2^{\ms{in}} \end{array} \!\! \right) . 
 \label{ddtmeans2}
\end{equation}
The input noise operators $x_T^{\ms{in}}$ and $p_T^{\ms{in}}$ describe the noise from the thermal bath,  $p_1^{\ms{in}} = -i(c_1^{\ms{in}} - c_1^{\ms{in}\dagger})$ describes the field entering through input 1, and $p_2^{\ms{in}} = -i(c_2^{\ms{in}} - c_2^{\ms{in}\dagger})$ that entering through input 2. The correlation functions of the thermal noise operators are given in Appendix A. 

The Langevin equations for the auxiliary cavity are 
\begin{equation}
\frac{d}{dt} \left( \!\! \begin{array}{c} X \\  P  \end{array} \!\! \right) = 
\left( \!\! \begin{array}{ccc}
- \frac{ \kappa}{2} & \Omega    \\
 -\Omega  &  - \frac{\kappa}{2}     
\end{array} \!\! \right) \!\! 
\left( \!\! \begin{array}{c}   X \\  P  \end{array} \!\! \right) + \sqrt{\kappa} \left( \!\! \begin{array}{c}   X_{\ms{in}} \\  P_{\ms{in}}  \end{array} \!\! \right) , 
\end{equation}
where the operators $X = a + a^\dagger$ and $P = -i(a - a^\dagger)$ are the amplitude and phase quadratures of the cavity mode, with $a$ the annihilation operator. The damping rate of the cavity is $\kappa$, and the input noise operators describe the single input. These operators are $X_{\ms{in}} = a^{\ms{in}} + a^{\ms{in}\dagger}$ and $P_{\ms{in}} = -i(a^{\ms{in}} - a^{\ms{in}\dagger})$ where $a^{\ms{in}}$ is a continuum annihilation operator with the same correlation functions as $c_1^{\ms{in}}$.  To connect the auxiliary input to output 1 of the oscillator we simply set 
\begin{equation} 
   a^{\ms{in}} =  - c_1^{\ms{out}} = - \sqrt{\tilde{k}} \tilde{x} + c_1^{\ms{in}}  \label{ch8links}
\end{equation}
where the minus sign accounts for the $\pi$ phase shift shown in Fig.~\ref{fig2}a. Similarly we connect the output of the auxiliary to input 2 by setting 
\begin{equation} 
   c_2^{\ms{in}} =   a^{\ms{out}} = \sqrt{\kappa} a - a^{\ms{in}}  \label{ch8links2} . 
\end{equation}
Substituting Eqs.(\ref{ch8links}) and (\ref{ch8links2}) into the equations of motion above for the oscillator and the cavity, the resulting coupled Langevin equations for the two systems are 
\begin{equation}
\frac{d}{dt} \left( \!\! \begin{array}{c}  \tilde{x}  \\  \tilde{p} \\ X \\ P \end{array} \!\! \right) = A  \left( \!\! \begin{array}{c}  \tilde{x}  \\  \tilde{p} \\ X \\ P \end{array} \!\! \right) +  \left(  \!\!\! \begin{array}{l}   \sqrt{\gamma} \, x_T^{\ms{in}}  \\   \sqrt{\gamma} \, p_T^{\ms{in}}  \\  \sqrt{\kappa} \, X_{\ms{in}} \\ \sqrt{\kappa} \, P_{\ms{in}}  \end{array} \!\!\! \right)
 \label{coup1}
\end{equation}
with 
\begin{equation}
 A = \left( \!\! \begin{array}{cccc}
- \frac{\gamma}{2} & \omega  &  0 & 0 \\
- \omega  & - \frac{\gamma}{2} \! & -\lambda &  0 \\ 
   0 &  0  & - \frac{\kappa}{2}  & \Omega  \\
   -\lambda & 0   & -\Omega & - \frac{\kappa}{2} 
 \end{array} \!\! \right)  ,  \label{cflang2}
\end{equation}
and we have defined $\lambda = \sqrt{8 \tilde{k} \kappa}$. 
The only noise driving the mechanical oscillator is now the thermal noise; the noise coming into input 2 from the auxiliary has cancelled the noise coming in input 1, which is a result of the $\pi$ phase shift applied to the auxiliary input. The auxiliary is driven by the noise from input 1 and is damped via the corresponding output channel at rate $\kappa$. It is this output that takes aways the entropy in the mechanical oscillator, since it is effectively damping to a thermal bath at zero temperature. The coupling between the two oscillators is given by the last terms on the RHS in the equations for $\dot{\tilde{p}}$ and $ \dot{X}$. These are the same as would be generated by an interaction Hamiltonian $H_{\ms{eff}} = \hbar \lambda \tilde{x} P/2$. Since the oscillation of the cavity mode continually transforms $P$ into $X$, we can replace $P$ with $X$ in this Hamiltonian without affecting the steady-state cooling, and this gives us a Hamiltonian equivalent to that in Eq.(\ref{hint}). If we now modulate the coupling strength $\lambda$ at the frequency difference between the two oscillators, then in the interaction picture the oscillators look as though they are resonant, and the result is the equations of motion for resolved sideband cooling. These equations are the same as those in Eq.(\ref{coup1}), but with $\Omega$ replaced with $\omega$.   
The modulation of the coupling strength can be realized by modulating the strength of the effective linear interaction between the mechanics and the transduction oscillators. Alternatively it can be achieved by imprinting a modulation on the fields that couple the auxiliary to the other components. 

\section{The performance of resolved-sideband cooling}
\label{cohfb}

Here we use the average number of phonons in the steady-state, $\bar{n} = \langle b^\dagger b \rangle$, to measure the degree of cooling.  To calculate this quantity for resolved-sideband cooling we solve the quantum Langevin equations (Eqs.(\ref{coup1}) and (\ref{cflang2}) with $\Omega$ replaced by $\omega$) in frequency space, and this gives us the spectrum of fluctuations of $\tilde{x}$ and $\tilde{p}$. Integrating these spectra over all frequencies gives us the steady-state expectation values of $\tilde{x}^2$ and $\tilde{p}^2$, which in turn gives us the average phonon number via  
\begin{equation}
    \bar{n} = \langle \tilde{x}^2 \rangle/4 + \langle \tilde{p}^2 \rangle/4 - 1/2 .  \label{bbrel}
\end{equation}
We derive the full expressions for $\langle \tilde{x}^2 \rangle$ and $\langle \tilde{p}^2 \rangle$ in Appendix A. These expressions are rather complex, but simplify greatly if we expand them to second-order in the small parameters given in Eq.(\ref{sbsmallp}).  Performing this expansion we find that  
\begin{eqnarray}
   \bar{n} &=& n_T \left( \frac{\gamma}{\kappa}  \right) \left[ 1 + \frac{\kappa^2}{\lambda^2} \right]  \label{bdbss} \\
     && -  \frac{n_T}{2}\left( \frac{\gamma}{\kappa} \right)^2  \left[ 1 + \frac{\kappa^2}{\lambda^2} + \frac{\kappa^4}{\lambda^4}  \right]  + \frac{1}{16}\left( \frac{\kappa}{\omega} \right)^2  + \frac{1}{8}\left( \frac{\lambda}{\omega} \right)^2 . \nonumber 
\end{eqnarray} 
Here the first line gives the dominant term, since it is first order while all terms on the second line are second order. Note that $\kappa/\lambda$ is not necessarily a small parameter; we will show it is of order $\varepsilon^{1/4}$ for optimal cooling. Note also that $\gamma n_T/\omega$ is second order in the small parameters. 

We note first that the dominant term in $\bar{n}$ gives the cooling performance as the ratio between the rate at which energy flows into the oscillator, $n_T \gamma$, to the maximum rate at which it can flow out of the auxiliary (and thus out of the oscillator), being $\kappa$, and tells us that that this rate is achieved when $\lambda \gg \kappa$. This makes sense, and if the dominant term were the only term determining $\bar{n}$ then we would get the best cooling by making $\lambda$ as large as possible. But this is not the case. The last two terms in $\bar{n}$ show that both $\kappa$ and $\lambda$ must be much smaller than $\omega$ to achieve ground-state cooling. This is because the linear $xX$ coupling between the oscillators only transfers energy efficiently between the two under the rotating-wave approximation, as is well-known~\cite{Tian08, Tian09, Wang11, Machnes12}.  The remaining term in $\bar{n}$ merely provides a correction to the dominant term, since it is second order in $\gamma/\kappa$, as long as $\lambda$ is not too much smaller than $\kappa$. Curiously it improves the cooling a little. 

The parameters $\gamma$ and $\omega$ are properties of the oscillator we want to cool, while $\lambda$ and $\kappa$ are parameters that we would ideally be able to choose as part of designing our controller. It is therefore natural to ask what values of $\lambda$ and $\kappa$ will give us the best cooling. It turns out that we can determine analytically the optimal value of $\lambda$ for a given value of $\kappa$ because this optimal value falls within the validity of our approximation. To do this we start by discarding the term proportional to $\kappa^4/\lambda^4$, an action that we will justify shortly. Differentiating the remaining terms with respect to $\lambda$, we find that the minimal value of $\bar{n}$ is reached when  
\begin{equation} 
    \lambda_{\ms{opt}} =  \sqrt{ \omega \sqrt{8\gamma \kappa n_T} \left(  1 - \frac{\gamma}{4\kappa} \right) } . 
\end{equation}
The assumption that we used in our expansion in powers of our small parameters was that $\lambda/\omega \sim \varepsilon$, but inspection shows that $\lambda_{\ms{opt}}/\omega \sim \varepsilon^{3/4}$, and so is lower than first-order by a factor of $(1/\varepsilon)^{1/4}$. This is not problematic unless higher-order terms in $(\lambda/\omega)$ that we have previously dropped (e.g. third and forth order terms) now have an order that is sufficiently low as to be near to the order of the leading-order terms from any of the other small parameters, such as $\gamma/\lambda$, which now has order $\varepsilon^{4/3}$. In that case we would have to include these high-order terms to be consistent. We will check the orders of the relevant terms below. We note now that with this value for $\lambda$, the term $\kappa^4/\lambda^4 \sim \varepsilon$, and so its total contribution to $\bar{n}$ is $\sim \varepsilon^3$. This is why we were justified in discarding it before we performed the minimization to obtain $\lambda_{\ms{opt}}$.

Substituting $\lambda_{\ms{opt}}$ into Eq.(\ref{bdbss}), and keeping terms only up to second order in $\varepsilon$, the minimum average phonon number is  
\begin{equation}
   \bar{n} = n_T  \frac{\gamma}{\kappa}   +  \sqrt{\frac{\gamma \kappa n_T}{2\omega^2}}   -  \frac{n_T }{2}\left( \frac{\gamma}{\kappa} \right)^2   +  \frac{\kappa^2}{16\omega^2}   . 
    \label{bdbssx}
\end{equation} 
The first term is the dominant term, proportional to $\varepsilon$, the second term is proportional to $\varepsilon^{1.5}$, and the last two terms are proportional to $\varepsilon^2$. The lowest-order term that we have discarded is $\mathcal{O}(\varepsilon^{2.5})$. 

The final step is to minimize $\bar{n}$ over $\kappa$ to obtain $\kappa_{\ms{opt}}$. In doing this we might assume that we can first discard the two second-order terms, because the two leading-order terms are sufficient to provide us with a minimum. However, upon doing this and substituting in the resulting optimal value for $\kappa$, we find that the fourth term contributes to the same order as the first and second. We therefore discard only the third term. Minimizing the remaining terms we obtain 
\begin{equation}
   \kappa_{\ms{opt}} =  d (\gamma n_T \omega^2)^{1/3} ,    \;\;\;\;\; d = [\sqrt{2}(\sqrt{5}-1)]^{2/3} , 
    \label{bdbssx2}
\end{equation} 
giving $\kappa_{\ms{opt}}/\omega \sim \varepsilon^{2/3}$. We now substitute this value for $\kappa$ into the expression for $\bar{n}$ above, and examine the order of the four terms. The first, second and fourth terms now all have order $\varepsilon^{4/3}$, and are thus all leading order. The third term can be discarded as it has order $\varepsilon^{8/3}$. We note also that we do not need to include any higher-order terms in $\lambda/\omega$ or $\kappa/\omega$ that we have previously discarded: since $\bar{n}$ is a symmetric function of $\lambda$ and $\kappa$, the lowest-order terms that we discarded are $(\lambda/\omega)^4 \sim \varepsilon^{3}$ and $(\kappa/\omega)^4 \sim \varepsilon^{2.66}$, and these are all significantly higher than the leading-order terms in $\bar{n}$. 

The minimal value of $\bar{n}$ is    
\begin{equation}
   \bar{n} = c \left( \frac{\gamma n_T}{\omega}  \right)^{2/3} =  c \left( \frac{n_T}{Q}  \right)^{2/3} . 
    \label{bdbssx3}
\end{equation} 
to leading order in our small parameters, with 
\begin{equation}
   c = \frac{1}{d} + \sqrt{\frac{d}{2}} + \frac{d^2}{16}  \approx 1.67 . 
    \label{bdbssx5}
\end{equation} 
The maximal cooling factor, defined as the ratio of the cooled phonon number, $\bar{n}$, to the initial phonon number $n_T$, is 
\begin{equation}
   R =  \frac{n_T}{\bar{n}} = c \, (n_T Q^2 )^{1/3} . 
    \label{bdbssx4} 
\end{equation}

The best possible cooling is not the only thing we wish to know. In order to compare with measurement-based cooling we would also like to know the best cooling that can be achieved for a given value of the output coupling rate $\tilde{k}$. To answer this question we substitute $\tilde{k}$ in for $\lambda$ in Eq.(\ref{bdbss}), which gives 
\begin{eqnarray}
    \bar{n}  & = & n_T \left( \frac{\gamma}{\kappa}  \left[ 1 + \frac{\kappa}{8 \tilde{k} } \right]  -  \frac{1}{2}\left( \frac{\gamma}{\kappa} \right)^2  \left[ 1 + \frac{\kappa}{8 \tilde{k} } + \frac{\kappa^2}{(8 \tilde{k})^2}  \right] \right)  \nonumber \\ 
    & & +  \frac{\kappa^2}{16\omega^2}  + \frac{\tilde{k} \kappa}{\omega^2} . 
    \label{bdbssk}
\end{eqnarray} 
Minimizing this expression exactly with respect to $\kappa$ gives a rather complex result, due to the need to solve a quartic equation. We can nevertheless obtain a simple expression that provides an upper bound on this minimum by choosing $\kappa$ so as to minimize the sum of the first and second-to-last terms only. The resulting value of $\kappa$ is
\begin{equation}
  \hat{\kappa} = 2 \left(  n_T \gamma \omega^2  \right)^{\! 1/3}, 
\end{equation}
so that $\gamma/\hat{\kappa} \sim \varepsilon^{4/3}$ and $\hat{\kappa}/\omega \sim \varepsilon^{2/3}$. Substituting this into Eq.(\ref{bdbssk}), and keeping terms up to order $\varepsilon^{5/3}$, we obtain 
\begin{eqnarray}
    \bar{n}  & \leq & \frac{n_T}{8}    \left( \frac{\gamma}{\tilde{k} } \right) + A  \left( \frac{\tilde{k}}{\omega}\right) + B  , 
    \label{bdbss3}
\end{eqnarray} 
with 
\begin{eqnarray}
    A & = & 2 \left( \frac{ n_T}{Q} \right)^{\! 1/3} \sim \varepsilon^{2/3} , \\
    B & = &  \frac{5}{8}  \left( \frac{n_T}{Q}\right)^{\! 2/3} \sim \varepsilon^{4/3} . \nonumber 
\end{eqnarray} 

\section{The performance of linear measurement-based cooling}
\label{measfb}

We now consider the best cooling that can be obtained by linear measurement-based feedback control, under which the dynamics of the system is described by Eq.(\ref{ch8mbfbcx}). Because the operator being measured is linear in the position and momentum of the oscillator, the oscillator is also linear, and the state of the system is always Gaussian, the dynamics is equivalent to that of a linear classical oscillator under a continuous measurement of position, driven by an additional white noise force that simulates exactly the quantum back-action of the measurement~\cite{DJ99, Jacobs14}. Because of this standard results from classical control theory can be applied to our system. The classical theory of linear optimal control, referred to as ``linear quadratic Gaussian'' (LQG) control tells us that if we wish to minimize a weighted sum of a quadratic function of the coordinates and a quadratic function of the control ``inputs'' (these inputs are the terms in the equations of motion for the momentum that come from the feedback force), then we should choose the feedback force to be a linear combination of the means of $x$ and $p$ given the observer's state of knowledge. However LQG theory does not apply directly to our problem, since we are interested in minimizing the energy without any particular reference to the control inputs. Nevertheless, the coherent feedback protocol we analyzed above is restricted, by the fact that the interaction is linear, to generating only linear dynamics in the system. It is reasonable therefore, in the interests of a fair comparison, that we also restrict the measurement-based feedback to generating linear dynamics. This means that the feedback force must be a linear combination of the expectation values of $\tilde{x}$ and $\tilde{p}$, and so can be written as 
\begin{equation}
   f(\delta,\Gamma) = - m \delta^2 \langle \tilde{x} \rangle  - \Gamma \langle \tilde{p} \rangle, 
\end{equation}
for two rate constants $\delta$ and $\Gamma$. The control inputs are then $-m \delta^2  \langle \tilde{x} \rangle$ and $-\Gamma \langle \tilde{p} \rangle$. 

In practice, and certainly in experiments today, the amount of force that can be applied induces motion on a timescale much slower than the oscillation of the resonator, and so we restrict ourselves to this regime here. The only effect of $\delta$ is to modify the frequency of the oscillator as $\omega' = \omega \sqrt{1 + \delta^2/\omega^2}$, and since $\delta \ll \omega$ in our regime, this has little effect on the dynamics. We can therefore drop $\delta$, leaving us with only one control parameter $\Gamma$. Clearly the larger $\Gamma$ the smaller will be the resulting steady-state energy, $\bar{n}$.  

Substituting the feedback force $f = - \Gamma \langle \tilde{p} \rangle$ into the master equation, Eq.(\ref{ch8mbfbcx}), the equations of motion for the means and variances of the real variables $x$ and $p$ are  
\begin{equation}
 \left( \!\! \begin{array}{c} d\langle x\rangle_{\ms{c}}  \\  d\langle p\rangle_{\ms{c}}  \end{array} \!\! \right) = 
\left( \!\! \begin{array}{ccc}
- \frac{ \gamma}{2} & 1/m     \\
 - m\omega^2  &  - \frac{ \gamma}{2}   - \Gamma  
\end{array} \!\! \right) \!\!
\left( \!\! \begin{array}{c}  \langle x\rangle_{\ms{c}}  \\  \langle p\rangle_{\ms{c}}  \end{array} \!\! \right) \! dt 
+ \sqrt{8\eta k} \! \left( \!\! \begin{array}{c} V_x \\  C \end{array} \!\! \right) \! dW
 \label{ch8mx1}
\end{equation}
and those for the variances are 
\begin{subequations}
\begin{eqnarray}
\dot{V}_x & = & \left(\frac{2}{m}\right) C - 8\eta k V^2_x - \gamma (V_x - V_x^T), \label{ddtcondVx} \\
\dot{C} & = & \left( \frac{1}{m} \right) V_p - m \omega^2 V_x - 8\eta k CV_x  - \gamma C , \\  
\dot{V}_p & = & -2m\omega^2 C - 8\eta k C^2 + 2k\hbar^2 - \gamma (V_p - V_p^T) ,   \;\;\label{ddtcondVp}
\label{cvhm}
\end{eqnarray}
\end{subequations}
where 
\begin{equation}
   C = \langle xp + px\rangle_{\ms{c}}/2 - \langle x\rangle_{\ms{c}} \langle p\rangle_{\ms{c}} 
\end{equation}
is the symmetrized  ``covariance'' of $x$ and $p$. It is important to remember that the means and variances in these equations are those of the state-of-knowledge of an observer who has access to the measurement results. We refer to them as the \textit{conditional} means and variances, which is why we denote the means with the subscript ``c''. 

The rate constant $\delta$ in the feedback force merely changes the effective frequency of the oscillator via $\omega \rightarrow \omega + \delta$, and so we have absorbed it into the definition of $\omega$. In practical situations, certainly those for nano-mechanical resonators, the feedback rate constants $\delta$ and $\Gamma$ are much smaller than $\omega$. The result of this is that $\delta$ will in fact have little effect on the cooling, but $\Gamma$ is very important as we will see below.  

Since the conditional means of $x$ and $p$ will be randomly fluctuating, the total variances averaged over all possible trajectories that the system may take while it is being controlled are given by adding the variances of the conditional means of $x$ and $p$ to the conditional variances. That is, if we denote the total variances by $\mathbb{V}_x$, $\mathbb{V}_p$, and $\mathbb{C}$, and the variances of the conditional means by $\mbox{\textbf{\textit{V}}}_{\bar{x}}$, $\mbox{\textbf{\textit{V}}}_{\bar{p}}$, and $\mbox{\textbf{\textit{C}}}$, then 
\begin{eqnarray}
\mathbb{V}_x & = & V_x +\mbox{\textbf{\textit{V}}}_{\bar{x}}  , \;\;\; \mathbb{V}_p = V_p +\mbox{\textbf{\textit{V}}}_{\bar{p}} , \;\;\; \mathbb{C} = C + \mbox{\textbf{\textit{C}}} . \;\;\;\; 
\label{cvhm2} 
\end{eqnarray}

We can derive the equations of motion of the variances of the conditional means by first using Ito calculus to derive the differential equations for $\langle x \rangle^2_{\ms{c}}$, $\langle p \rangle^2_{\ms{c}}$, and $\langle x \rangle_{\ms{c}} \langle p\rangle_{\ms{c}}$ from Eq. (\ref{ch8mx1}). Taking averages on both sides of the differential equations for these square means gives us the differential equations for the second moments of the means. From these we can obtain the equations of motion for the variances of the means, and these are 
\begin{equation}
\frac{d}{dt}\! \left( \!\! \begin{array}{c} \tilde{\mbox{\textbf{\textit{V}}}}_{\bar{x}} \\ \tilde{\mbox{\textbf{\textit{V}}}}_{\bar{p}} \\  \tilde{\mbox{\textbf{\textit{C}}}} \end{array} \!\! \right) = - \!
\left( \!\! \begin{array}{ccc}
 \gamma & 0  & -2\omega   \\ 
 0 & \gamma + 2\Gamma  &  2\omega \\ 
\omega  & -\omega   &   \gamma + \Gamma 
\end{array} \!\! \right) \!\! 
 \left( \!\! \begin{array}{c} \tilde{\mbox{\textbf{\textit{V}}}}_{\bar{x}} \\ \tilde{\mbox{\textbf{\textit{V}}}}_{\bar{p}} \\  \tilde{\mbox{\textbf{\textit{C}}}} \end{array} \!\! \right) \, + \,  8k \! \left( \!\! \begin{array}{c} ( \tilde{V}_x )^2 \\ ( \tilde{C} )^2 \\  \tilde{C} \tilde{V}_x \end{array} \!\! \right) 
 \label{ddtVsofmeans}
\end{equation}
Here we have written the equations in terms of dimensionless (scaled) versions of the variances, defined by 
\begin{eqnarray}
   \tilde{\mbox{\textbf{\textit{V}}}}_x & \equiv &  \frac{\mbox{\textbf{\textit{V}}}_x}{(\Delta x)^2} , \;\;\;\; \tilde{\mbox{\textbf{\textit{V}}}}_p  \equiv   \frac{\mbox{\textbf{\textit{V}}}_p}{(\Delta p)^2} , \;\;\;\;  \tilde{\mbox{\textbf{\textit{C}}}} \equiv \frac{\mbox{\textbf{\textit{C}}}}{\Delta x \Delta p}  .  \label{scaledVs}
\end{eqnarray}
These scaled variances are those of the scaled variables $\tilde{x}$ and $\tilde{p}$. Using the scaled variances simplifies the equations, and exposes the important rate constants in the dynamics. From now on any variance with a tilde will indicate the dimensionless version of that variance (e.g. $\tilde{V}_x \equiv V_x/(\Delta x)^2$). The scaled versions of the thermal variances are 
\begin{eqnarray}
   \tilde{V}_x^T =  \tilde{V}_p^T  = (1 + 2 n_T ) \equiv \tilde{V}^T . 
\end{eqnarray}
The harmonic oscillator ground state has $\tilde{V}_x = \tilde{V}_p = 1$.  

To calculate the total variances in the steady-state we need to determine the steady-states of both the conditional variances and the variances of the means. This can be done by setting the left-hand-sides of the equations of motion to zero, and solving the resulting algebraic equations. There is a big difference between the differential equations for the conditional variances and those for the variances of the means: we have written the equations for the latter in matrix form because they are linear, whereas the equations for the former are not, and as they stand do not have analytic solutions. 

If the harmonic oscillator had no damping, so that $\gamma$ were zero, there would be an analytic solution for the steady-states of the conditional variances, being   
\begin{equation}
  \tilde{V}_x^0 = \frac{\sqrt{2}}{\sqrt{\eta (\xi + 1) }}, \;\;\;  \tilde{V}_p^0  = \xi \tilde{V}_x^0, \;\;\; \tilde{C}^0  = \frac{\sqrt{ \xi - 1}}{\sqrt{ \eta(\xi + 1)}} .  
  \label{sscondVs}
\end{equation}
where 
\begin{equation} 
  \xi =  \sqrt{ 1 + \eta r^2} , \;\;\;\;  r = \frac{8\tilde{k}}{\omega} .
  \label{ch8defxi}
\end{equation}
We see from these solutions that if we want to keep the oscillator close to the ground state, for which $\tilde{V}_x = \tilde{V}_p = 1$, then $\xi$ must be close to unity, and thus $r^2 \ll 1$ (or equivalently $(8\tilde{k})^2 \ll \omega^2$, assuming that $\eta \sim 1$).  

While we cannot obtain an analytic solution for the steady-states of the conditional variances for all values of $\gamma$, we can obtain an approximate solution valid when $\gamma n_T$ is much smaller than $\tilde{k}$ and $\omega$. Since $\tilde{k}$ is also much smaller than $\omega$ there is more than one way to do this expansion. We do it by using the solution above for $\gamma = 0$ as our zeroth-order solution, and writing $\gamma = q \tilde{k}$ and $\gamma = c q \omega$, where $q$ is the small parameter, and $c = \tilde{k}/\omega$ is unrestricted. We then solve to obtain the steady-states to first-order in $q$. We are subsequently free to expand the zeroth-order solutions to second order in $c = \varepsilon_6 = \tilde{k}/\omega \propto r$ to obtain solutions to second order in $\varepsilon$. The result of the first expansion is  
\begin{eqnarray} 
\tilde{V}_x^{\ms{ss}} & = &  \tilde{V}_x^0 + \left(\frac{\gamma}{8\eta \tilde{k}}\right) \left( \frac{1 + \eta r \tilde{V}_x^0/2 }{1 + \eta r \tilde{V}_x^0} \right) \frac{(\tilde{V}^T - \tilde{V}_x^0)}{\tilde{V}_x^0} ,   \nonumber \\ 
\tilde{C}^{\ms{ss}} & = & \tilde{C}^0 + \left(\frac{\gamma}{2\omega}\right) \frac{(\tilde{V}^T - \tilde{V}_x^0)}{1  + \eta r \tilde{V}_x^0} ,   \nonumber \\ 
\tilde{V}_p^{\ms{ss}} & = & \tilde{V}_p^0 + \left( 1 + \eta r \tilde{C}^0 \right) (\tilde{V}_x^{\ms{ss}} -\tilde{V}_x^0)  \nonumber \\ 
  & &  + \eta r \tilde{V}_x^0 ( \tilde{C}^{\ms{ss}} - \tilde{C}^0) .   \nonumber  
\end{eqnarray} 
Now expanding to second order in $r \propto \varepsilon_6$ we obtain   
\begin{eqnarray} 
\tilde{V}_x^{\ms{ss}} & = &  \tilde{V}_x^0 + \left(\!\frac{\gamma}{8\eta \tilde{k}}\!\right)\!\! \left(\! 1 - \frac{\sqrt{\eta}  r}{2} \!\right) \!\! \left(\! \sqrt{\eta} \left[1+ \frac{\eta r^2}{8}\right]\! \tilde{V}^T -1 \!\right) ,    \nonumber \\ 
\tilde{C}^{\ms{ss}} & = & \tilde{C}^0 + \left(\frac{\gamma}{2\omega}\right) \left(1 -  \sqrt{\eta} r \right) \left(\tilde{V}^T - \frac{[1- \eta r^2/8]}{\sqrt{\eta}} \right) ,   \nonumber \\ 
\tilde{V}_p^{\ms{ss}} & = & \tilde{V}_p^0 + (\tilde{V}_x^{\ms{ss}} -\tilde{V}_x^0) + \sqrt{\eta} r  ( \tilde{C}^{\ms{ss}} - \tilde{C}^0)  .   \nonumber
\end{eqnarray} 
For efficient detection these equations simplify considerably, and we see more clearly the effects of the measurement and thermal noise:  
\begin{eqnarray} 
\tilde{V}_x^{\ms{ss}} & = &  \tilde{V}_x^0 +  \left(\frac{\gamma}{4 \tilde{k}}\right) \left[ n_T  \left(1 - \frac{4\tilde{k}}{\omega} \right)  + ( 2 n_T + 1) \left( \frac{2\tilde{k}}{\omega} \right)^{\!\! 2} \right]   ,   \nonumber \\ 
\tilde{C}^{\ms{ss}} & = & \tilde{C}^0 + \left(\frac{\gamma}{\omega}\right) \left[n_T  \left(1 -  \frac{8\tilde{k}}{\omega} \right)  + \left( \frac{2\tilde{k}}{\omega} \right)^{\!\! 2} \right] ,  \nonumber \\ 
\tilde{V}_p^{\ms{ss}} & = & \tilde{V}_p^0 + (\tilde{V}_x^{\ms{ss}} -\tilde{V}_x^0) +  r  ( \tilde{C}^{\ms{ss}} - \tilde{C}^0)  .   \nonumber
\end{eqnarray} 
Since we are expanding to first-order in $\gamma/\tilde{k}$ and second order in $r$, we should drop terms proportional to $(\gamma/\tilde{k}) r^2$ as they contribute no more than the other third-order terms that have already been dropped. We have kept these in the above equations merely to show how the second-order terms in $r$ affect the solution. 

Calculating the steady-state variances of the means is straightforward because the equations of motion are linear, although the resulting expressions are rather cumbersome. We find that these variances will only be small, and thus the oscillator close to the ground state, when $\Gamma \gg \tilde{k}$. This makes sense because the noise from the measurement that causes the means to fluctuate is proportional to $\tilde{k}$, and it is the job of the feedback damping at rate $\Gamma$ to counteract it. We therefore expand the solutions for the variances of the means in the small parameter $\tilde{k}/\Gamma$. If we want to allow $\Gamma$ to be smaller than $\omega$ then we can assume that $\tilde{k}/\Gamma \sim \sqrt{\varepsilon}$. We keep only first-order terms in $\tilde{k}/\Gamma$ but we do not drop any terms in $\Gamma/\omega$, so $\Gamma$ is not restricted to being small compared to $\omega$. The resulting expressions for the variances of the means are  
\begin{eqnarray} 
 \tilde{\mbox{\textbf{\textit{V}}}}_{\bar{x}}^{\ms{ss}} & = & \frac{4\tilde{k}}{\Gamma} \left[  \left(\! 1 + \frac{\Gamma^2}{\omega^2} \! \right) \! (\tilde{V}_x^{\ms{ss}})^2 + (\tilde{C}^{\ms{ss}})^2  + 2 \left(\frac{\Gamma}{\omega}\right) \tilde{C}^{\ms{ss}}\tilde{V}_x^{\ms{ss}}  \right]  ,   \nonumber  \\ 
\tilde{\mbox{\textbf{\textit{V}}}}_{\bar{p}}^{\ms{ss}} & = & \frac{4\tilde{k}}{\Gamma} \left[ (\tilde{V}_x^{\ms{ss}})^2 + (\tilde{C}^{\ms{ss}})^2   \right] ,   \nonumber  \\
\tilde{\mbox{\textbf{\textit{C}}}}^{\ms{ss}} & = & -\frac{4\tilde{k}}{\omega} (\tilde{V}_x^{\ms{ss}})^2 .    \nonumber 
\end{eqnarray} 
We can see from these expressions that we cannot achieve good cooling if we make $\Gamma$ too large. This is because our feedback force damps only the momentum, and so to confine the position as well as the momentum we need the oscillation of the oscillator to transform position into momentum (and vice versa) on a timescale at least as fast as the damping rate $\Gamma$. 

Now we have the steady-state solutions for the conditional variances and the variances of the means, we can combine them to obtain the total variances as per Eq.(\ref{cvhm2}). Since $\langle \tilde{x} \rangle = \langle \tilde{p} \rangle = 0$, Eq.(\ref{bbrel}) tells us that $\langle b^\dagger b \rangle = \tilde{\mathbb{V}}_x/4 + \tilde{\mathbb{V}}_p/4 - 1/2$, which we can use to obtain the mean steady-state phonon number. To second order in $\varepsilon$ the result is  
\begin{eqnarray} 
    \bar{n} & = &   n_T  \left(\frac{\gamma}{8 \tilde{k}} - \frac{\gamma}{2\omega} \right)  + \frac{4\tilde{k}^2}{\omega^2}  \nonumber \\  & & + \frac{2\tilde{k}}{\Gamma} \left( 1 + \frac{\Gamma^2}{2\omega^2} \right) \left( 1 + n_T \frac{\gamma}{4 \tilde{k}} \right) . 
    \label{nbarm}
\end{eqnarray}
The answer depends both on $\Gamma$ and $\tilde{k}$. It would be nice to find the optimal value of $\Gamma$ and thus eliminate $\Gamma$ from the expression. However this optimal value is $\Gamma = \sqrt{2}\omega$, which is unrealistic for practical purposes, and would also imply that the feedback can significantly change the frequency $\omega$. In this case we would also be able to increase $\omega$ to reduce $\bar{n}$ further. 

So instead of minimizing with respect to $\Gamma$, we assume $\tilde{k} \ll \Gamma \ll \omega$, and thus $\tilde{k}/\Gamma \sim \Gamma/\omega \sim \sqrt{\varepsilon}$. Keeping only terms up to first-order in $\varepsilon$ we then have 
\begin{equation}
    \bar{n} =  \frac{n_T}{8} \left( \frac{\gamma}{\tilde{k}} \right) + 2 \left( \frac{\tilde{k}}{\Gamma} \right) . 
    \label{mfbdbk}
\end{equation} 
Finally, we minimize this expression with respect to the measurement strength to determine the best possible cooling. The optimal measurement strength is 
\begin{equation}
   \tilde{k}_{\ms{opt}} =   \frac{1}{4}  \sqrt{n_T \gamma \Gamma} , 
\end{equation} 
and the best cooling is 
\begin{equation}
   \bar{n} =  \left(\sqrt{2} + \frac{1}{2} \right) \sqrt{n_T\left(\frac{\gamma}{\Gamma}\right)} . 
   \label{nmbf}
\end{equation}       
The cooling factor is 
\begin{equation}
   R \equiv \frac{n_T}{\bar{n}} =  \frac{1}{\left(\sqrt{2} + \frac{1}{2} \right)} \sqrt{n_T\left(\frac{\Gamma}{\gamma}\right)} . 
\end{equation}

\section{Comparing measurement-based and coherent feedback cooling}
\label{seccomp}

We begin by examining the expressions for $\bar{n}$ for sideband cooling and measurement-based feedback for a fixed output coupling rate $\tilde{k}$, given respectively by Eqs.(\ref{bdbssk}) and (\ref{nbarm}). In writing these equations now, we keep only the terms that make the most important contribution to limiting the cooling. We have 
\begin{align} 
    \bar{n} & \approx   n_T  \left(\frac{\gamma}{8 \tilde{k}}\right)  + \frac{(8\tilde{k})^2}{16\omega^2} + \frac{2\tilde{k}}{\Gamma} 
       & \mbox{(meas. feedback)}  \label{mbfx} \\
    \bar{n}  & \approx n_T \left(\frac{\gamma}{8 \tilde{k}} +  \frac{\gamma}{\kappa}\right) +  \frac{\kappa^2 + 2 (8\tilde{k}) \kappa}{16\omega^2}  & \mbox{(sideband cooling)}  
\end{align}
The origins of the various terms in these expressions can be clearly identified. The terms proportional to $\gamma$ give the value of $\bar{n}$ that results from the balance between the rate at which energy is injected into the oscillator from the bath at the rate $n_T\gamma$, and the rate at which it is extracted by the controller. We can read off the energy extraction rates as 
\begin{align} 
  R & = 8\tilde{k} & \mbox{(meas. feedback)}  \\
  R  & = \frac{(8\tilde{k})\kappa}{(8\tilde{k})+\kappa} & \mbox{(sideband cooling)} 
\end{align}
The rate for sideband cooling is the series combination of two conductances, which makes intuitive sense. Curiously the measurement has the advantage as far as the extraction rate is concerned, but this is because this rate takes into account only the desirable effect of the purification induced by the measurement. This purification comes at the expense of projection noise, to which we will return below. 

The second set of terms, those proportional to $1/\omega^2$ are remarkably similar for the two controllers. The term in the expression for measurement-based feedback comes purely from the squeezing of the conditional momentum variance. It is due to the fact that the reduction in the position variance due to the measurement of position causes an increase in the momentum variance, which is precisely the back-action noise of the measurement. This term is not a fundamental restriction of measurement-base cooling, it is present only because we are restricted to a linear interaction with the resonator, and thus a measurement that is linear in the coordinates $x$ and $p$. 

For sideband cooling the term proportional to $8\tilde{k}\kappa = \lambda^2$ is due to the breakdown of the rotating-wave approximation; when $\lambda$ is small compared to the oscillator frequency the $xX$ interaction acts purely to transfer energy between the two, but this is no longer true as $\lambda$ is increased relative to $\omega$, in which case it generates excitations in both systems. This limitation on the coherent feedback cooling is not a fundamental one, but is due to the linearity of the interaction. The term proportional to $\kappa^2$ is due to the fact that the damping of the auxiliary interferes with the energy transfer process. Quantum mechanically this can be attributed to the quantum Zeno effect, since the damping is a measurement process that inhibits the unitary dynamics. Since the joint system is linear, and is therefore equivalent to a noisy classical system, there must also be a classical interpretation. One possibility is that in changing the transfer function of the auxiliary, the damping inhibits the energy transfer in a way that is similar to taking the auxiliary off-resonance with the oscillator. The only way to avoid this limitation on the cooling appears to be to make the coherent control process time-dependent, rather than using an auxiliary with a constant Hamiltonian. We will return to this topic below. 

So far the terms in the expressions for both cooling schemes parallel each other to a large extent. While they may have somewhat different origins they have very similar forms, and would lead to similar cooling behavior. For example, the heating due to the back-action noise of the measurement is similar to the heating due to the correction to the rotating-wave approximation that appears in sideband cooling, and both are caused by the nature of the interaction. The final term in the expression for measurement-based cooling is quite different, as it has no parallel in sideband cooling. It is the heating due to the noise on the mean position and momentum that comes from the random nature of the measurement results, and is a necessary companion to the purification generated by the measurement. This noise is sometimes referred to as the \textit{projection noise} of the measurement. This noise, being proportional to $\tilde{k}$, is the projection noise of the position measurement when the oscillator is in its ground state. It is the role of the feedback force to counteract this noise, which is why the resulting heating is proportional to $\tilde{k}/\Gamma$. This noise is not a fundamental limitation on measurement-based cooling, but is due to the fact that it is the position of the oscillator that is measured.  

It is the heating due to the projection noise that makes measurement-based feedback significantly inferior to sideband cooling for cooling an oscillator via a linear interaction. There are two reasons for this. The first is that the heating terms in sideband cooling have $\omega$ on the bottom line, whereas the heating due to the projection noise is suppressed only by $\Gamma$. In practical situations, and certainly in current experiments, $\Gamma$ is considerably smaller than $\omega$. The second reason is that the heating term coming from the projection noise is first-order in $\tilde{k}$. The heating term in sideband cooling that is proportional to $\tilde{k}$ also has a factor of $\kappa/\omega \ll 1$. Since $\kappa$ and $\tilde{k}$ can be expected to be similar for optimal cooling, the heating for sideband cooling is effectively second-order in $\tilde{k}/\omega$. Because of this, even if we set $\Gamma \sim \omega$, and thus replace $\gamma/\Gamma$ in Eq.(\ref{nmbf}) with $Q$, the maximal cooling for measurement-based feedback scales as $1/\sqrt{Q}$, while that for sideband cooling scales as $1/Q^{2/3}$. 

What would happen if we were able to apply a classical feedback ``force'' to damp the position as well as the momentum? In this case the term $\Gamma^2/(2\omega^2)$ would no-longer appear in Eq.(\ref{nbarm}) and we would no-longer need $\Gamma \lesssim \omega$. In the limit in which $\Gamma\rightarrow\infty$ the projection noise would be eliminated, and the performance of measurement-based cooling would be 
\begin{equation}
  \bar{n} = 1.5 \left( \frac{n_T}{Q} \right)^{2/3} 
\end{equation}
which is obtained by keeping only the first two terms in Eq.(\ref{mbfx}). This is slightly better than that for sideband cooling, but requires quite different interactions and very large feedback forces. 
 
We have found that the factors that place limits on both coherent and measurement-based feedback for cooling an oscillator with a linear interaction are not fundamental restrictions imposed by quantum mechanics, but are due to the linear nature of the interaction. Both control methods could perform much better with a non-linear interaction. Nevertheless, resolved-sideband cooling is able to make better use of the linear interaction and achieve much better cooling that measurement-based feedback. We have found that it is not the back-action noise of the measurement which leads to this difference in performance, but the projection noise of the measurement. 

Finally, there is another important difference between coherent feedback and measurement-based feedback in this linear cooling scenario. The performance of the coherent scheme can be greatly improved even without a nonlinear coupling, merely by making the interaction rate $\lambda$ time-dependent~\cite{Wang11, Machnes12}. This eliminates the need for the rotating-wave approximation, with the result that the energy in the oscillator can be swapped into the auxiliary within a single period of the oscillator. The maximal cooling is then 
\begin{equation}
  \bar{n} \sim \frac{n_T}{Q}  . 
\end{equation}
Measurement-based feedback cannot be improved in this way, and instead requires a non-linear interaction. 

\section{Acknowledgements:} 
KJ and FS were partially supported by the NSF under project Nos.\ PHY-1005571 and PHY-1212413, and KJ was partially supported by the ARO MURI grant W911NF-11-1-0268. HN and MJ were supported by the Australian Research Council, and MJ was also supported by the Air Force Office of Scientific Research (AFOSR) under grant AFOSR FA2386-09-1-4089 AOARD 094089. 

\vspace{5mm}

\appendix

\section{Describing measurement-based feedback using the input-output formalism}
\label{quantumprob}

To analyze measurement-based feedback in Section~\ref{measfb} we used the stochastic master equation, while we used the quantum noise formalism of input-output theory to analyze coherent feedback. In the standard formalism of quantum mechanics used by physicists, the approach used to derive the former is very different from the analysis that leads to the latter. Since the derivation of the quantum Langevin equations by Collett and Gardiner (CG) involves approximations, it is not at all clear that they describe the same physical process as the stochastic master equation. Nevertheless, one can show explicitly that the auto-correlation functions of the output fields of the former agree exactly with those of the measurement records of the latter, and this is enough to show equivalence for most applications. But doing so is not simple (see for example \citep[pp.\ 470-474]{Jacobs14}). 

There is another way to formulate measurement theory in quantum mechanics, which uses measure theory in the way that it is used in probability theory. The resulting structure is called \textit{quantum probability}~\cite{Chang15}. This formulation of measurement theory can be used to construct both the quantum noise formalism and continuous measurement theory, and in this case it is clear by construction that the two descriptions refer to the same process. This quantum probability formulation of input-output theory was first developed by Hudson and Parthasarathy (HP)~\cite{Hudson84, Parthasarathy14}, and exploited for continuous measurement by Belavkin \cite{VPB92a, BHJ07}.

Because the HP formalism contains an explicit mapping between the quantum noise operators and the classical measurement record --- the latter being a classical stochastic process --- it allows us to write a measurement-based feedback process using quantum Langevin equations, something that is not possible in the input-output formalism as derived by CG using the standard formulation of quantum mechanics~\cite{Gardiner85}. To do this for the measurement-based feedback cooling scheme described in Section~\ref{physimp} we first write down the quantum Langevin equations for the oscillator, which are given by Eq.(\ref{ddtmeans1}). The output field that our controller measures is 
\begin{equation}
  x_1^{\ms{out}} = c_1^{\ms{out}} + c_1^{\ms{out}\dagger} = - x_1^{\ms{in}} + \sqrt{k} x . 
  \label{x1out}
\end{equation}
The HP formalism now goes beyond the CG formalism by telling us that the white noise quantum field $x_1^{\ms{out}}$ can be interpreted immediately, without any further machinery, as a classical white noise process. That is, we can fully describe the stream of measurement results from a homodyne detection performed on the field $x_1^{\ms{out}}$ by $x_1^{\ms{out}}$ \textit{itself}. The reason for this is that, in the quantum probability framework, $x_1^{\ms{out}}$ is a classical noise process; its quantum nature is captured by the fact that it does not commute with other noise processed that are also contained in the full probability space of events. 

The quantum nature of the output field $x_1^{\ms{out}}$ is important: it means that we cannot treat \textit{both} $x_1^{\ms{out}}$ and $p_1^{\ms{out}}$ as classical noise sources. This is because when we measure the output field we cannot choose to measure both in the $x$-basis and $p$-basis at the same time. We could chose a measurement that gave us partial information about both $x$ and $p$, but it would produce a stream of measurement results that was neither equal to $x_1^{\ms{out}}$ or $p_1^{\ms{out}}$. Practically what this means is that we are free to send both $x_1^{\ms{out}}$ and $p_1^{\ms{out}}$ into a quantum system in order to process them, but we can only send one of them through a classical processing device. The dynamics of all quantum systems will preserve the correct relationship between non-commuting operators, but classical processing will in general not do so because it is less restricted. 

Interpreting $x_1^{\ms{out}}$ now as the classical measurement record, we can obtain our estimates $\langle x\rangle_{\ms{c}}$ and $\langle p\rangle_{\ms{c}}$ from $x_1^{\ms{out}}$ in the usual way by using Eqs.(\ref{ch8mx1}) and setting  
\begin{equation}
   dW = dx_1^{\ms{out}} -  \sqrt{8\eta k} \langle x\rangle_{\ms{c}} dt .  
\end{equation}
To complete the feedback loop we include the feedback force $-\Gamma \langle p\rangle_{\ms{c}}$ in the quantum Langevin equations for the mechanical oscillator, which are then
\begin{equation}
\frac{d}{dt} \left( \!\! \begin{array}{c}  \tilde{x}  \\  \tilde{p}  \end{array} \!\! \right) = 
\left( \!\! \begin{array}{ccc}
- \frac{ \gamma}{2} & \omega    \\
 -\omega  &  - \frac{ \gamma}{2} 
\end{array} \!\! \right) \!\! 
\left( \!\! \begin{array}{c}  \tilde{x}  \\  \tilde{p}  \end{array} \!\! \right) -  \left( \!\! \begin{array}{c}  0 \\  \Gamma \langle p\rangle_{\ms{c}}  \end{array} \!\! \right)  + \mathbf{v}_{\ms{in}}  . 
 \label{ddtmeans1x}
\end{equation}
While it may seem odd that the c-number $\langle p\rangle_{\ms{c}}$ now appears in a differential equation for operators, as usual any c-number merely acts as a multiple of the identity operator. 

The measurement-based feedback process is now described by the coupled equations (\ref{ch8mx1}), (\ref{ddtcondVx}) -- (\ref{ddtcondVp}), and (\ref{ddtmeans1x}). These equations can be compared more easily to the Langevin equations describing the coherent feedback protocol than can the SME. We can further eliminate the output field from Eqs.(\ref{ch8mx1}) and (\ref{ddtmeans1x}) by using Eq.(\ref{x1out}). The result is a set of Langevin equations driven by the input noise operators. As with the Langevin equations for sideband cooling, we can use these to calculate power spectra and correlation functions, thus providing an alternative method for analyzing measurement-based feedback protocols. Examples of the use of the HP input-output theory to describe measurement feedback can be found in~\cite{Naoki14}. 

\onecolumngrid

\section{Calculating the exact steady-state for sideband cooling}
\label{appsecsslin}

Steady-states for linear open quantum systems can be obtained by solving the Langevin equations in the frequency domain, and then integrating the spectrum over all frequencies. This integration can be done with an integral formula that can be found in Gradshteyn and Ryzhik, and which we give below. To begin we recall that the Langevin equations for the coupled oscillators, when the interaction is modulated at the frequency $\Omega-\omega$, is given by 
\begin{equation}
   \dot{\mathbf{x}} = M \mathbf{x} + \boldsymbol{\xi}(t)  \;\;\;\;\; \mbox{with} \;\;\;\;\; 
 \mathbf{x} = \! \left( \!\! \begin{array}{l} \tilde{x}  \\ \tilde{p} \\ X \\ P   \end{array} \!\!\! \right) \! , \; 
 \boldsymbol{\xi} = \! \left(  \!\!\! \begin{array}{l}   \sqrt{2\Gamma} \, x_T^{\ms{in}}  \\   \sqrt{2\Gamma} \, p_T^{\ms{in}}  \\  \sqrt{2K} \, X_{\ms{in}} \\ \sqrt{2K} \, P_{\ms{in}}  \end{array} \!\!\! \right) \! , \;
 M \! = \! \left( \!\! \begin{array}{cccc}
- \Gamma & \omega  &  0 & 0 \\
- \omega  & - \Gamma \! & -\lambda &  0 \\ 
   0 &  0  & - K & \omega  \\
   -\lambda & 0   & -\omega & - K
 \end{array} \!\! \right) \! ,   \nonumber 
\end{equation}
and for compactness we have defined $\Gamma = \gamma/2$ and $K = \kappa/2$. 

To solve the equations of motion for $\mathbf{x}$ in the frequency domain we take the Fourier transform of both sides of the equation. Denoting the frequency space variables with a caret, e.g., 
\begin{equation}
   \hat{\mathbf{x}}(\nu)  = \frac{1}{\sqrt{2\pi}} \int_{-\infty}^\infty  \mathbf{x}(t) e^{-i v t} dt , 
\end{equation}
the equations of motion become $- i\nu \hat{\mathbf{x}} = M \hat{\mathbf{x}} + \hat{\boldsymbol{\xi}}(\nu)$. Rearranging gives 
\begin{equation}
  \hat{\mathbf{x}} = - (M + i\nu I)^{-1} \hat{\boldsymbol{\xi}}(\nu) \equiv A(\nu) \hat{\boldsymbol{\xi}}(\nu) .  \label{eq132}
\end{equation}
The dynamical variables are therefore given by a linear combination of the noise sources, where the coefficients are functions of $\nu$ and therefore filter the noise. Inverting the matrix $M + i\nu I$, for which an algebraic software package is invaluable, we obtain the matrix 
\begin{equation} 
 A(\nu) = \frac{1}{D(\nu)}\left( \! \begin{array}{cccc}
f(\Gamma)  g(K)  & \omega g(K) & \omega\lambda f(K) & \omega^2 \lambda \\
- \omega g(K) - \lambda^2 \omega  & f(\Gamma)  g(K)  &  \lambda f(\Gamma) f(K)&   \omega \lambda f(\Gamma)  \\
 \omega \lambda f(\Gamma) &\omega^2  \lambda & f(K)  g(\Gamma) & \omega g(\Gamma)  \\
 \lambda f(\Gamma) f(K) & \omega \lambda f(K) & - \omega g(\Gamma) - \lambda^2 \omega & f(K)  g(\Gamma) 
 \end{array} \! \right) \! , 
\end{equation}
with 
\begin{eqnarray}
   f(\alpha) & = & \alpha - i\nu , \;\;\;\;\;\; g(\alpha)  =   (i\omega + i \nu - \alpha)(i\omega - i \nu + \alpha) , 
\end{eqnarray}
and 
\begin{eqnarray}
   D(\nu) =  \left[ f(K)^2 + \omega^2 \right] \left[ f(\Gamma)^2 + \omega^2 \right] - \lambda^2 \omega^2.  
\end{eqnarray}
Two important properties of the matrix $A$ are i) that each element is a ratio of polynomials, and ii) that the imaginary unit $i$ and the frequency $\nu$ always appear together in Eq.(\ref{eq132}). This second property means that taking the complex conjugate of any element of $A$ is the same as replacing $\nu$ with $-\nu$. 

The steady-state variance of a dynamical variable is given by integrating the spectrum for that variable over all $\nu$. The spectrum for $\tilde{x}$ (for example) is given by 
\begin{equation}
 S_x(\nu) = F(\nu,-\nu)    \;\;\;\; \mbox{where}  \;\;\;\;   \langle \hat{\tilde{x}}(\nu) \hat{\tilde{x}} (\nu') \rangle = F(\nu, \nu') \delta(\nu + \nu') . 
\end{equation}
We can obtain the correlation functions for the dynamical variables $\hat{\mathbf{x}}(\nu)$ directly from those of the noise sources: 
\begin{equation}
    \langle \hat{\mathbf{x}}(\nu) \hat{\mathbf{x}} (\nu') \rangle = A(\nu) \langle \hat{\boldsymbol{\xi}}(\nu) \hat{\boldsymbol{\xi}}(\nu')^{\ms{T}} \rangle A(\nu')^{\ms{T}} = A(\nu) G A(\nu')^{\ms{T}} \delta(\nu + \nu') , 
\end{equation}
where $G$ is the correlation matrix for the noise sources, and is given by 
\begin{equation}
    G = 2 \left( \! \begin{array}{cccc}
\Gamma (2 n_T + 1) & 0  &  0 & 0 \\
0 & \Gamma (2 n_T + 1) & 0  & 0 \\ 
0 & 0  &  K & 0 \\
0 & 0 & 0  & K 
 \end{array} \! \right) \! . 
\end{equation}
The spectrum for $\tilde{x}$ is 
\begin{equation}
    S_x(\nu) = \frac{2\Gamma\left( 2n_T+1 \right) |g(K)|^2 \left( |f(\Gamma)|^2 +\omega^2 \right) + 2K(\lambda \omega)^2 \left( |f(K)|^2 +\omega^2 \right)}{D(\nu)D(-\nu)}   
\end{equation}
and that for $\tilde{p}$ is 
\begin{eqnarray}
    S_p(\nu) & = & \frac{\Gamma\left( 2n_T+1 \right) \left\{ |g(K)|^2 \left( |f(\Gamma)|^2 +\omega^2 \right)  + 2 \lambda^2 \omega^2 \mbox{Re}[g(K)] +  (\lambda^2 \omega)^2 \right\} }{D(\nu)D(-\nu)}  
    + \frac{K(\lambda \omega)^2 |f(\Gamma)|^2 \left( |f(K)|^2 +\omega^2 \right)}{D(\nu)D(-\nu)}   \;\;\;
\end{eqnarray}

The expressions for the spectra contain $8^{\ms{\textit{th}}}$-order polynomials in the denominator. If these polynomials had no special structure, they would likely be impossible to integrate analytically. The fact that this is possible is due to the following remarkable integral formula, which is a slightly simplified version of a formula in Gradshteyn and Ryzhik~\cite{Gradshteyn94}: 
\vspace{1.5ex}
\begin{equation}
  I_n \equiv  \int_{-\infty}^{\infty} \frac{y_n(\nu)}{z_n(\nu)z_n(-\nu)} d\nu =   \frac{\pi }{a_0} \left| \frac{M_n}{L_n} \right| , 
\end{equation}
where $z(\nu)$ must satisfy $z(-\nu) = z^*(\nu)$, 
\begin{eqnarray}
   y(\nu) & = & b_0 \nu^{2n-2} + b_1 \nu^{2n-4} + \cdots + b_{n-1}  , \\
   z(\nu) & = & a_0 \nu^n + a_1 \nu^{n-1} + \cdots + a_{n}  ,  
\end{eqnarray} 
and 
\begin{equation}
      M_n = \left|
\begin{array}{ccccc}
 b_0  & b_1  & b_2  & \ldots  & b_{n-1}  \\
 a_0 &  a_2 & a_4 & \ldots & 0    \\
0 &  a_1 & a_3 & \ldots & 0    \\
 \vdots & \vdots  & \vdots & \vdots & \vdots  \\   
  0 &  0 & 0 & \ldots & a_n 	
\end{array}
\right|
 ,   \;\;\;\;\;   
     L_n = \left|
\begin{array}{ccccc}
  a_1 &  a_3 & a_5 & \ldots & 0  \\
 a_0 &  a_2 & a_4 & \ldots & 0    \\
0 &  a_1 & a_3 & \ldots & 0    \\
 \vdots & \vdots  & \vdots & \vdots & \vdots  \\   
  0 &  0 & 0 & \ldots & a_n  
\end{array}
\right| . 
\end{equation}
Note that $L_n$ and $M_n$ are determinants of matrices that differ only by their first row.  
\vspace{1.5ex}

Here we need the case $n=4$, for which the integral is 
\begin{eqnarray}
  I_4 = \pi \left| \frac{(b_0/a_0)( a_2 a_3- a_1 a_4 ) - b_1 a_3 + b_2 a_1  + (b_3/a_4)(a_0 a_3 - a_1 a_2)}{a_0 a_3^2 + a_1^2 a_4 - a_1 a_2 a_3} \right| .    
\end{eqnarray} 
Using the above integral formula, and the fact that the steady-state mean squares of $\tilde{x}$ and $\tilde{p}$ are 
 \begin{equation}
   \langle \tilde{x}^2 \rangle_{\ms{ss}} = \frac{1}{2\pi} \int_{-\infty}^{\infty} S_x(\nu) d\nu  , \;\;\;\;\;\;\; \langle \tilde{p}^2 \rangle_{\ms{ss}} = \frac{1}{2\pi} \int_{-\infty}^{\infty} S_p(\nu) d\nu , 
\end{equation}
we obtain 
\begin{align}
   \langle \tilde{x}^2 \rangle_{\ms{ss}} & = \frac{(2 n_T + 1) A_1 + A_2}{F} + \frac{(2 n_T + 1) B_1 + B_2}{C F} , \\
   \langle \tilde{p}^2 \rangle_{\ms{ss}} & =  \frac{(2 n_T + 1) G_1+ G_2}{F} + \frac{(2 n_T + 1) J_1 + J_2}{CF} . 
\end{align}
Here we have defined 
\begin{eqnarray}
   A_1 & = & r \Gamma \left( b^2 + a (b - 4 \omega^2) + \lambda^2 \omega^2  \right) + 4 \Gamma \left( 2 r^2 + b \right) \left( a K + b \Gamma \right)  \\
    A_2 & = & r K \lambda^2 \omega^2  \\
    B_1 & = & r \Gamma a b^2 ( r^2 + \Gamma K + \lambda^2 \omega^2 )   \\
    B_2 & = & r \Gamma b \lambda^2 \omega^2 ( r^2 + \Gamma K + \lambda^2 \omega^2 )  \\
    G_1 & = & r \Gamma \left( b^2 + a (2c - b)  (r + 2 K + a/r) ( a K  + b\Gamma) + 3 \omega^2 \lambda^2 \right) \\
    G_2 & = & \lambda^2 K \left( a K + b \Gamma + r ( a + c + \omega^2) \right)   \\
    J_1 & = & r \Gamma (r^2 + \Gamma K + \omega^2) (\lambda^4 \omega^2 + a b^2 - 2 b \lambda^2 \omega^2)   \\
     J_2 & = & r b  \lambda^2 \Gamma^2  K (r^2 + \Gamma K + \omega^2) 
 \\
     F & = & 2 r^2 [ \Gamma K (r^2 + 4\omega^2) -  \lambda^2 \omega^2 ] \\
     C & = & ab - \lambda^2 \omega^2 , 
\end{eqnarray}
with 
\begin{eqnarray}
   a & = &  \Gamma^2 + \omega^2, \;\;\;\; b = K^2 + \omega^2, \;\;\;\; c =  K^2 - \omega^2, \;\;\;\; r = \Gamma + K . 
\end{eqnarray}
\twocolumngrid


%

\end{document}